\documentclass[aps,preprint,superscriptaddress,tightenlines]{revtex4}
\usepackage{graphicx} 
\usepackage{dcolumn}  

\def\figwid{2.9in}

\graphicspath{{ps}}

\input{defns}
\begin{document}


\preprint{\vbox{ \hbox{   }
                 \hbox{BELLE-CONF-0416 }
                 \hbox{ICHEP04 11-0661 } 
}}

\title{ \quad\\[0.5cm]  Improved Measurements of Color-Suppressed Decays \newline {\boldmath $\Bzerobar \to \Dzero \pizero $}, {\boldmath $\Dzero \eta$ }  and
{\boldmath $\Dzero \omega$ } }

\affiliation{Aomori University, Aomori}
\affiliation{Budker Institute of Nuclear Physics, Novosibirsk}
\affiliation{Chiba University, Chiba}
\affiliation{Chonnam National University, Kwangju}
\affiliation{Chuo University, Tokyo}
\affiliation{University of Cincinnati, Cincinnati, Ohio 45221}
\affiliation{University of Frankfurt, Frankfurt}
\affiliation{Gyeongsang National University, Chinju}
\affiliation{University of Hawaii, Honolulu, Hawaii 96822}
\affiliation{High Energy Accelerator Research Organization (KEK), Tsukuba}
\affiliation{Hiroshima Institute of Technology, Hiroshima}
\affiliation{Institute of High Energy Physics, Chinese Academy of Sciences, Beijing}
\affiliation{Institute of High Energy Physics, Vienna}
\affiliation{Institute for Theoretical and Experimental Physics, Moscow}
\affiliation{J. Stefan Institute, Ljubljana}
\affiliation{Kanagawa University, Yokohama}
\affiliation{Korea University, Seoul}
\affiliation{Kyoto University, Kyoto}
\affiliation{Kyungpook National University, Taegu}
\affiliation{Swiss Federal Institute of Technology of Lausanne, EPFL, Lausanne}
\affiliation{University of Ljubljana, Ljubljana}
\affiliation{University of Maribor, Maribor}
\affiliation{University of Melbourne, Victoria}
\affiliation{Nagoya University, Nagoya}
\affiliation{Nara Women's University, Nara}
\affiliation{National Central University, Chung-li}
\affiliation{National Kaohsiung Normal University, Kaohsiung}
\affiliation{National United University, Miao Li}
\affiliation{Department of Physics, National Taiwan University, Taipei}
\affiliation{H. Niewodniczanski Institute of Nuclear Physics, Krakow}
\affiliation{Nihon Dental College, Niigata}
\affiliation{Niigata University, Niigata}
\affiliation{Osaka City University, Osaka}
\affiliation{Osaka University, Osaka}
\affiliation{Panjab University, Chandigarh}
\affiliation{Peking University, Beijing}
\affiliation{Princeton University, Princeton, New Jersey 08545}
\affiliation{RIKEN BNL Research Center, Upton, New York 11973}
\affiliation{Saga University, Saga}
\affiliation{University of Science and Technology of China, Hefei}
\affiliation{Seoul National University, Seoul}
\affiliation{Sungkyunkwan University, Suwon}
\affiliation{University of Sydney, Sydney NSW}
\affiliation{Tata Institute of Fundamental Research, Bombay}
\affiliation{Toho University, Funabashi}
\affiliation{Tohoku Gakuin University, Tagajo}
\affiliation{Tohoku University, Sendai}
\affiliation{Department of Physics, University of Tokyo, Tokyo}
\affiliation{Tokyo Institute of Technology, Tokyo}
\affiliation{Tokyo Metropolitan University, Tokyo}
\affiliation{Tokyo University of Agriculture and Technology, Tokyo}
\affiliation{Toyama National College of Maritime Technology, Toyama}
\affiliation{University of Tsukuba, Tsukuba}
\affiliation{Utkal University, Bhubaneswer}
\affiliation{Virginia Polytechnic Institute and State University, Blacksburg, Virginia 24061}
\affiliation{Yonsei University, Seoul}
  \author{K.~Abe}\affiliation{High Energy Accelerator Research Organization (KEK), Tsukuba} 
  \author{K.~Abe}\affiliation{Tohoku Gakuin University, Tagajo} 
  \author{N.~Abe}\affiliation{Tokyo Institute of Technology, Tokyo} 
  \author{I.~Adachi}\affiliation{High Energy Accelerator Research Organization (KEK), Tsukuba} 
  \author{H.~Aihara}\affiliation{Department of Physics, University of Tokyo, Tokyo} 
  \author{M.~Akatsu}\affiliation{Nagoya University, Nagoya} 
  \author{Y.~Asano}\affiliation{University of Tsukuba, Tsukuba} 
  \author{T.~Aso}\affiliation{Toyama National College of Maritime Technology, Toyama} 
  \author{V.~Aulchenko}\affiliation{Budker Institute of Nuclear Physics, Novosibirsk} 
  \author{T.~Aushev}\affiliation{Institute for Theoretical and Experimental Physics, Moscow} 
% ...
\collaboration{The Belle Collaboration}

\noaffiliation

\begin{abstract}
We present preliminary improved measurements of the branching fractions of the color-suppressed  
decays $\Bzerobar \to  \Dzero \hzero$  
where $\hzero$ 
represents the three light neutral mesons $\pizero$, $\eta$ and $\omega$.
The measurements are based on a data sample of 140 \ensuremath{ \mathrm{fb}^{-1}\,} collected at the 
\ensuremath{ \Upsilon(4S) }\, with the Belle detector at
the KEKB energy-asymmetric \ensuremath{e^+e^-}\, collider, corresponding to seven times
the luminosity of the previous Belle measurements.  
\end{abstract}


\maketitle


{\renewcommand{\thefootnote}{\fnsymbol{footnote}}}
\setcounter{footnote}{0}

\section{Introduction}

\begin{figure}[htb]
\begin{center}
\begin{minipage}{4.1in}
\includegraphics[width=2in]{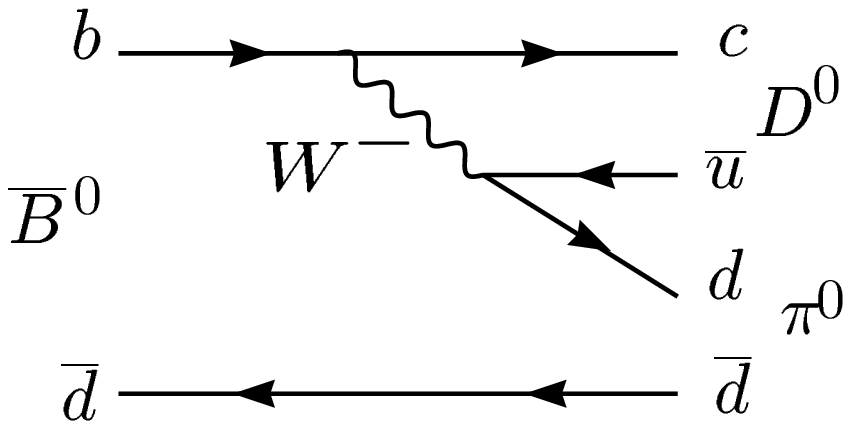}
\includegraphics[width=2in]{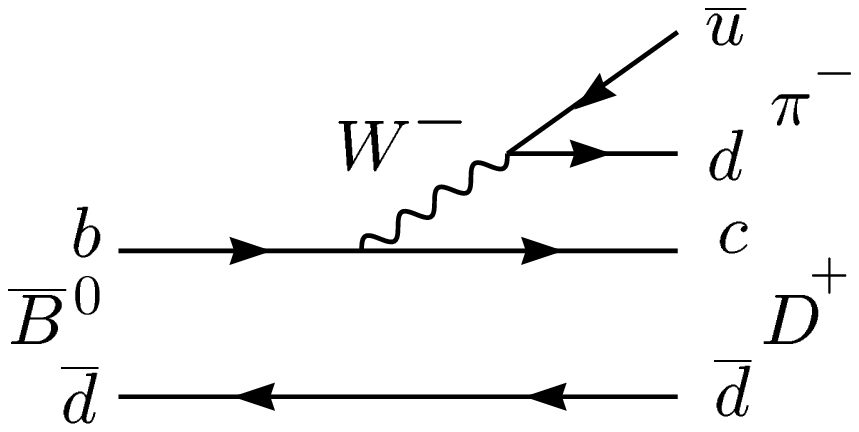}
\end{minipage}
\medskip
\caption{Tree level internal (left) and external (right) spectator 
diagrams for $\overline{B}\to D\pi$ decays.}
\label{dpi-feynman}
\end{center}
\end{figure}

The weak decays $\Bzerobar \ra \Dstze \hzero$~\cite{CC}, where $\hzero$ represents a light neutral 
meson, are expected to proceed predominantly through internal spectator diagrams, as illustrated
in Fig.~\ref{dpi-feynman}.
The color matching requirement between the quarks from the virtual $\Wmi$ and the
other quark pair results in these decays being ``color-suppressed'' relative to decays
such as $\Bzerobar \ra \Dstpl \hmi$, which proceed through external spectator diagrams.

Previous measurements of 
$\Bzerobar$ decays into $\Dstze \pizero$, $\Dzero \eta$, $\Dzero \omega$, and $\Dzero \rho^0$ by the Belle collaboration~\cite{ref:Belle,ref:Belle2},
and of $\Bzerobar$ into $\Dstze \pizero$ by the CLEO collaboration~\cite{ref:CLEO}, 
and of $\Bzerobar$ decays into $\Dstze \pizero$, $\Dstze \eta$, $\Dstze \omega$, $\Dzero \eta'$
by the BaBar collaboration~\cite{ref:Babar} indicate color suppressed branching
fractions in the approximate range $(2$--$4)\times 10^{-4}$.
This is substantially in excess of theoretical expectations from ``naive" 
factorization models~\cite{ref:Beneke,ref:NeuSte,ref:NeuPet,ref:Chua,ref:Rosner,ref:Deandrea,ref:ChRos} in the range $(0.3$--$1.7) \times 10^{-4}$. 
 
Several approaches to achieving a better theoretical description~\cite{ref:NeuPet,ref:Chua,ref:SCET,ref:pQCD} have been developed. 
They extend upon the factorization approach with consideration of 
final state interactions and consequent simultaneous treatment of  isospin amplitudes of  color-suppressed and color-allowed decays.
The possibility that similar effects could have dramatic implications
on the measurement potential of direct $CP$ violation asymmetries in charmless decays, together with some degree of 
discrepancy between the prior Belle~\cite{ref:Belle} and BaBar~\cite{ref:Babar} measurements provide strong 
motivation for more precise measurements of the color-suppressed decays.

In this paper we report improved branching fraction measurements 
of  $\Bzerobar$ decays into $\Dzero \pizero$, 
$\Dzero \eta$, $\Dzero \omega$. The measurements are  
based on a $140~{\rm fb}^{-1}$ data sample, which contains 152 million $B\overline{B}$ pairs, 
collected  with the Belle detector at the KEKB asymmetric-energy
$e^+e^-$ (3.5 on 8~GeV) collider~\cite{KEKB}
operating at the $\Upsilon(4S)$ resonance.  This corresponds to seven times the luminosity  
of the previous Belle measurements~\cite{ref:Belle} and almost twice that of the 
earlier BaBar measurements~\cite{ref:Babar}.

The Belle detector is a large-solid-angle magnetic
spectrometer that consists of a three-layer silicon vertex detector (SVD),
a 50-layer central drift chamber (CDC), an array of
aerogel threshold \v{C}erenkov counters (ACC), 
a barrel-like arrangement of time-of-flight
scintillation counters (TOF), and an electromagnetic calorimeter
comprised of CsI(Tl) crystals (ECL) located inside 
a super-conducting solenoid coil that provides a 1.5~T
magnetic field.  An iron flux-return located outside of
the coil is instrumented to detect $K_L^0$ mesons and to identify
muons (KLM).  The detector
is described in detail elsewhere~\cite{Belle}.

\section{Event Selection}

Color-suppressed $\Bzerobar$ meson decays are reconstructed 
from candidate $\Dzero$ mesons that are combined with  
light neutral meson candidates $\hzero$. The $\Dzero$ mesons are reconstructed
in three decay modes: $\Kpi$, $\Ktwopi$, and $\Kthreepi$ while the light neutral mesons  $\hzero$ are
reconstructed in the decay modes: $\pizero \to \gamgam$, $\eta \to \gamgam$, $\eta \to \threepi$ 
and $\omega \to \threepi$. The invariant masses at each stage of the decay chains are required to
be consistent within $2.5-3 \sigma$ mass resolution or natural width windows around the 
nominal masses of the assumed particle types. 
Vertex and mass constrained fits are performed for decays with charged products such as
the  three $\Dzero$ decays  and $\eta \to \threepi$; mass constrained fits  are performed on 
the $\pizero \to \gamgam$ and $\eta \to \gamgam$ candidates; and vertex constrained fits are performed on $\omega \to \threepi$ candidates  due to the large natural width of the $\omega$ meson. These 
kinematic fits result in improved energy and momenta of the candidate mesons. 
 
Charged tracks are required to have impact parameters within $\pm 5 \cm$ of the interaction
point along the positron beam axis and within $1 \cm$ in the transverse plane. Each track is identified
as a kaon or pion according to a likelihood ratio derived from the responses of the TOF and ACC
systems and energy loss measurements from the CDC. The likelihood ratio is required to exceed
0.6 for  kaon candidates. This requirement is $88\%$ efficient for kaons with a misidentification
rate for pions of $8.5\%$. 

The photon pairs that constitute $\pizero$ candidates are required to have energies greater than
50 MeV and an invariant mass within a $\pm 3\sigma$ ($\sigma = 5.4 \MeVcsq$) mass window around the
nominal $\pizero$ mass.

Candidate $\eta$ mesons that decay to $\gamgam$ are required to have photon energies $E_{\gamma}$  greater than 
$100 \MeV$. In addition the energy asymmetry 
$\frac{|E_{\gamma_1}-E_{\gamma_2}|}{E_{\gamma_1}+E_{\gamma_2}}$,
is required to be less than 0.9. The $\eta$ candidates are required to have invariant masses 
within $2.5\sigma$ mass windows of the nominal mass, where $\sigma = 10.6\MeVcsq$ for 
the $\eta \to \gamgam$ mode and $3.4 \MeVcsq$ for the $\eta \to \threepi$ mode.   
If the photons that comprise the $\eta \to \gamgam$ candidate are found to contribute to
any $\pizero \to \gamgam$ the candidate is excluded. 
The $\pizero$ decay products 
of the $\eta \to \threepi$ and $\omega \to \threepi$ candidates 
are required to have CM momentum greater than $200$ and $500 \MeV$, respectively.
The $\omega$ candidates are required to have invariant masses within $\pm3 \Gamma$ of the nominal mass value, 
where $\Gamma$ is the natural width of $8.9 \MeVcsq$.

Invariant masses of the $\Dzero$ candidates are required to be within $\pm 2\sigma$ of the nominal
mass where $\sigma$ is $8, 12$ and $5 \MeVcsq$ for the $\Kpi$, $\Ktwopi$, and $\Kthreepi$ modes respectively.
The CM momentum of the $\pizero$ in the $\Ktwopi$ mode is required to be greater than $400 \MeV$.

\section{ {\boldmath \B}  reconstruction }

The $\Bzerobar$ candidates are reconstructed from combinations of  $\Dzero$ and $\hzero$ using the
improved energy and momenta resulting from the vertex and mass constrained fits.  

Two kinematic variables are used to distinguish signal candidates from backgrounds: the
 beam-energy constrained mass 
 $\Mb = \sqrt{  (\Eb)^2 -  |\sum \vec{p}_{i}^{*}|^2 )}$ and energy difference 
 $\De = \sum E_{i}^{*}  - \Eb $ 
where $\Eb$ is the CM energy, and 
$E_{i}^{*} $ , $\vec{p}_{i}^{*} $ are the CM energy and momenta, respectively, which are summed over 
the $\Dzero$ and $\hzero$ meson decay candidates.

The resolution of $\Mb$ is approximately $3 \MeVcsq$ for all modes,  dominated by the  
beam energy spread, whereas the $\De$ resolution varies substantially among modes
depending particularly on the number of $\pizero$ in the final state.    
Candidates within the broad region $|\De| < 0.25 \GeV $ and $  5.2 \GeVcsq < \Mb < 5.3 \GeVcsq$
are selected for further consideration. Where more than one candidate is found in a single
event the one with the smaller $\sum \chi_{i}^{2} / N_{i}  $ is chosen, where $\chi_{i}^{2}$ and 
the number of degrees of freedom $N_{i}$  are obtained from the
the kinematic fits to the  $\Dzero$ and $\hzero$.

 
A common $\Mb$ signal region of  $  5.27 \GeVcsq < \Mb < 5.29 \GeVcsq$ is used for all final states. The signal
region definitions in $\De$ are mode dependent with  $|\De| < 0.05 \GeV $ for $\omega \to \threepi$
and $\eta \to \threepi$ modes ,  and  $|\De| < 0.08 \GeV $ for $\pizero \to \gamgam$ and $\eta \to \gamgam$   modes. 
The event yields and efficiencies presented in the following sections correspond to these signal regions.

\section{Continuum Suppression}

At energies close to the $\Upsilon(4S)$ resonance the
production cross section of $\epem \to \qqbar$ $( q = u,d,s,c )$ is approximately three times that of $\BB$ production, 
making continuum background suppression essential in all modes. 
The jet-like nature of the continuum events allows 
event shape variables to discriminate between them and the more spherical $\BB$ events.

The discrimination power of seven event shape variables is combined into a single Fisher discriminant~\cite{fw} whose variables
include the angle between the thrust axis
of the \B candidate and the thrust axis of the rest of the event 
($\cos{\theta_{T}}$), the sphericity variable, and five modified Fox-Wolfram moments~\cite{fw}. 

Monte Carlo event samples of continuum $\qqbar$ events and signal events for each of the 
final states considered are used to construct probability density functions (PDFs) for the 
Fisher discriminant~\cite{fw} and $\cos{\theta_{B}}$, where $\theta_{B}$ is the angle between
the \B  flight direction and the beam direction in the $\Upsilon(4S)$ rest frame. 
The products of the PDFs for these two variables give signal and 
continuum likelihoods ${\cal L}_{s}$ and ${\cal L}_{\qqbar}$ for each 
candidate, allowing a selection to be applied to the likelihood ratio 
${\cal L} = {\cal L}_{s} / ({\cal L}_{s} + {\cal L}_{\qqbar} )$.    


Monte Carlo studies of the signal significance $N_{s}/\sqrt{N_{s}+N_{b}}$, where $N_s$ and $N_b$ are signal and
background yields (using signal branching fractions from previous measurements), as a function of a cut on the
likelihood ratio ${\cal L}$ indicate a rather smooth behavior. Although the optimum significance is generally in the
range 0.6-0.7, a looser cut of ${\cal L} > 0.5 $  is applied for all modes in order to reduce systematic uncertainties.

For the  $\Bzerobar \to \Dzero\omega$ mode the polarized nature of the $\omega$ allows additional discrimination against
backgrounds to be achieved with an additional requirement of $|\cos{\theta_{hel}}|>0.3$, where the
helicity angle $\theta_{hel}$  is defined as the 
angle between the \B  flight direction in the $\omega$ rest frame and the vector perpendicular to the $\omega$
decay plane in the $\omega$ rest frame.

\section{Backgrounds from other {\boldmath \B} decays}

\begin{figure}[htb]
\begin{center}
\begin{minipage}{6.1in}
\includegraphics[width=\figwid]{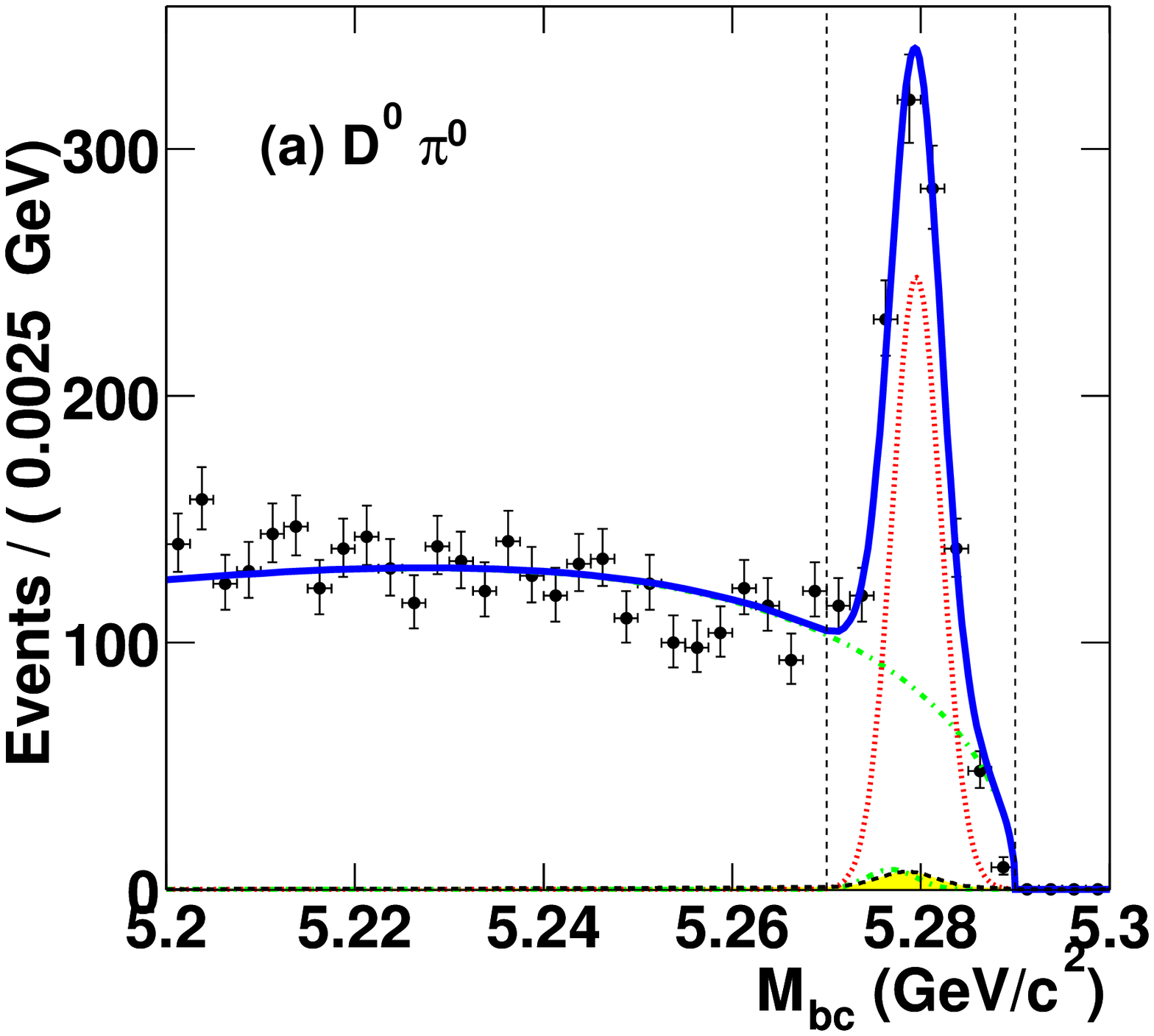}
\includegraphics[width=\figwid]{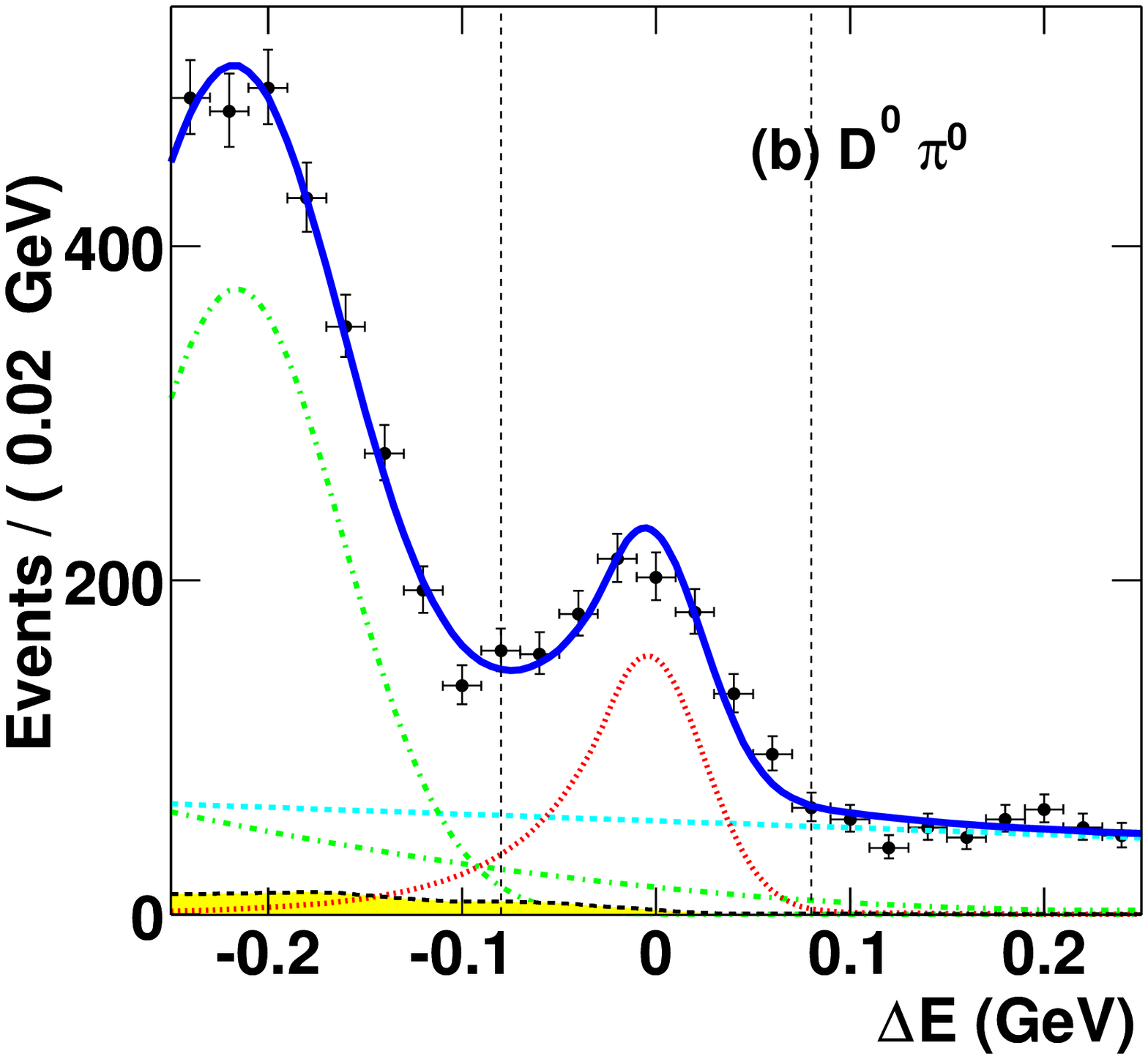}
\end{minipage}
\medskip
\caption{
Distributions of  (a) $\Mb$ and  (b) $\De$ for $\Bzerobar \to \Dzero \pizero$. Points with error bars 
represent the data, 
the solid line shows the result of the fit and the dotted line represents the signal contribution. 
The crossfeed contributions are represented by the shaded areas.
The vertical dashed lines represent the signal regions.
For (a) the dashed-dotted line shows the continuum-like background contribution, with peaking background contributions represented
by the small dashed line.
For (b) the dashed line shows the continuum background contribution, the 
dashed-dotted lines show the $B$ background components contribution.
}
\label{fig-pi0}
\end{center}
\end{figure}

\begin{figure}[htb]
\begin{center}
\begin{minipage}{6.1in}
\includegraphics[width=\figwid]{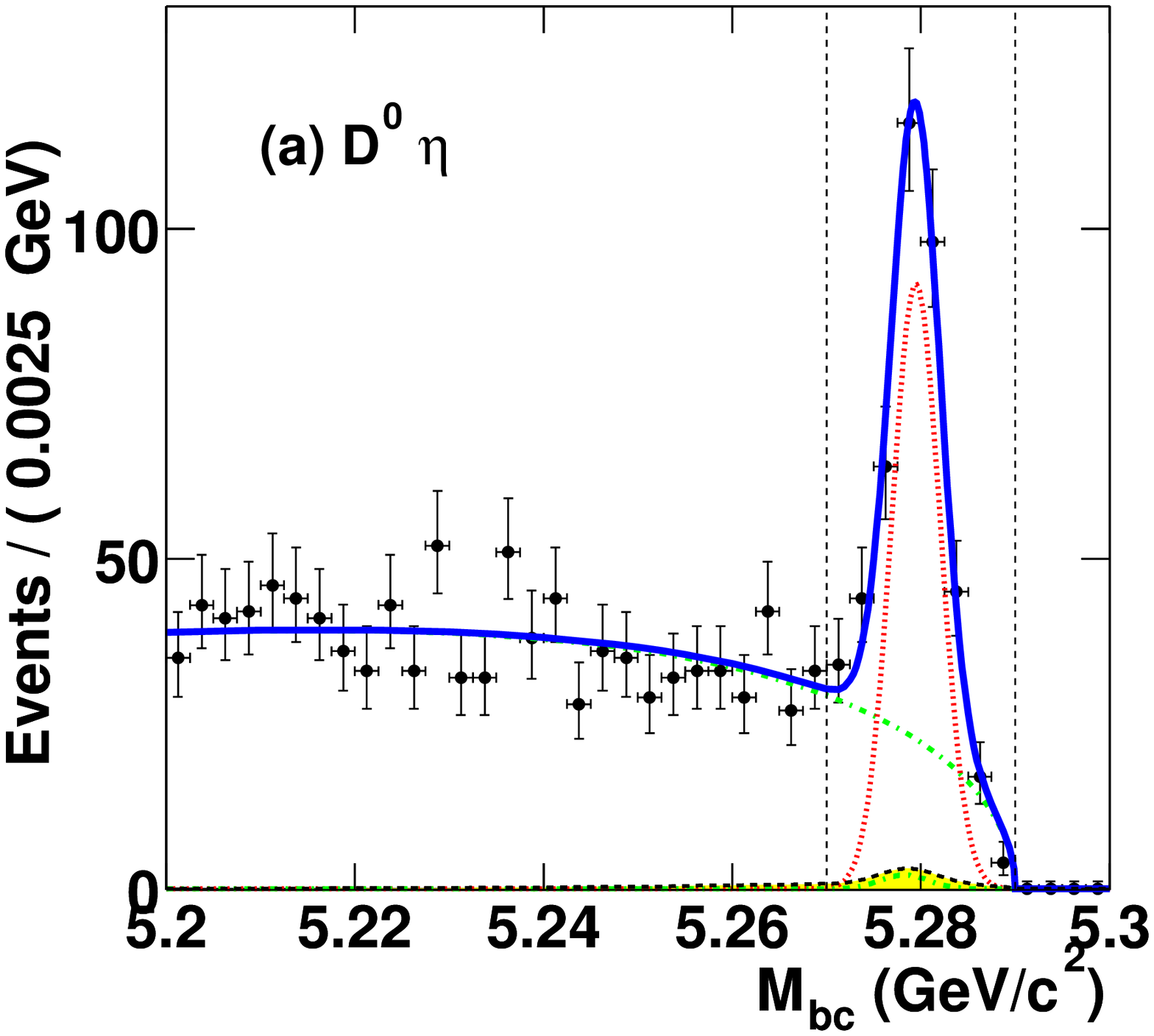}
\includegraphics[width=\figwid]{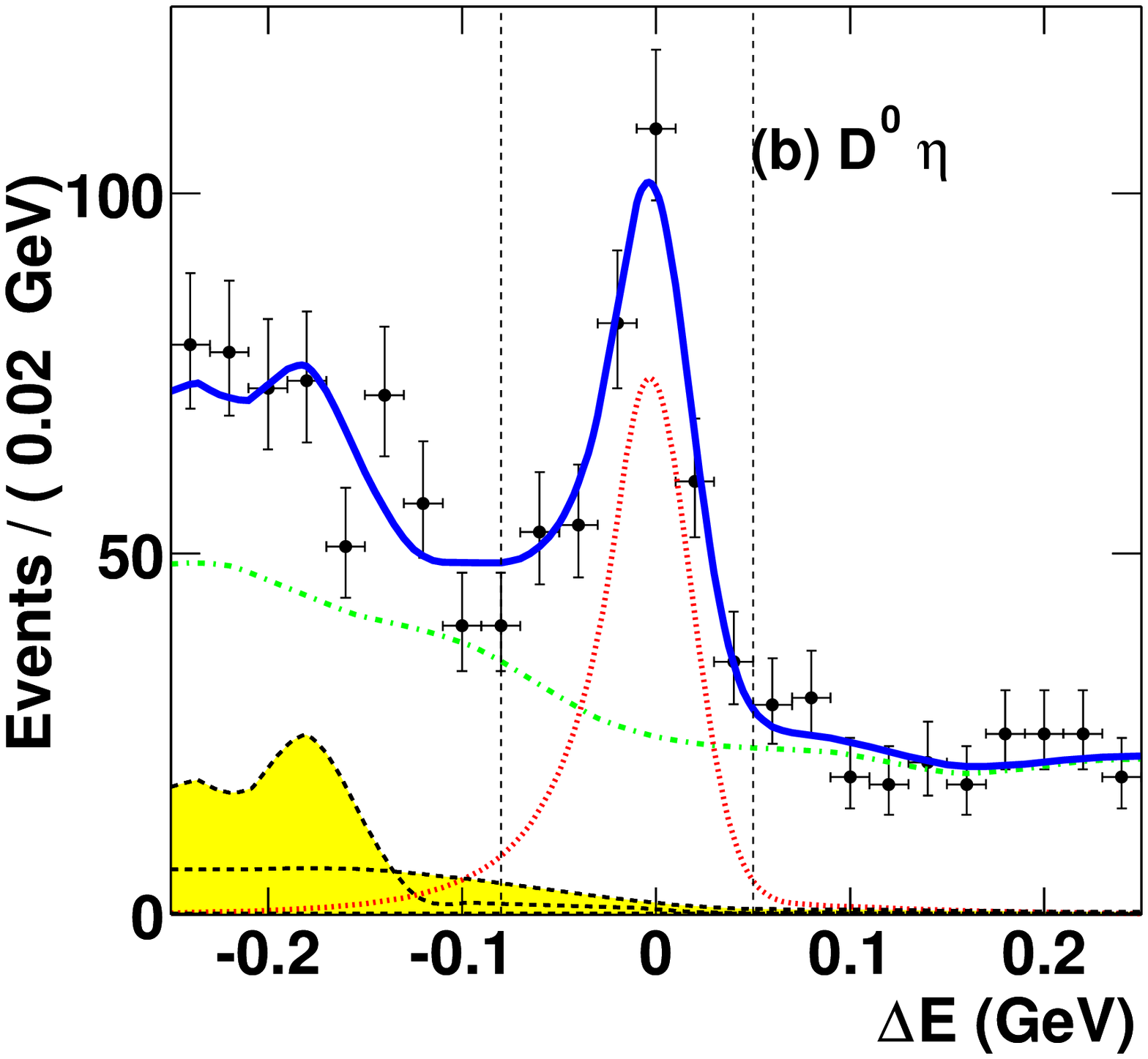}
\end{minipage}
\medskip
\caption{
Distributions of  (a) $\Mb$ and  (b) $\De$ for $\Bzerobar \to \Dzero \eta$ . Points with error bars 
represent the data, 
the solid line shows the result of the fit and the dotted line represents the signal contribution.
The crossfeed contributions are represented by the shaded areas.
The vertical dashed lines represent the signal region.
For (a) the dashed-dotted line shows the continuum-like background contribution, with peaking background contributions
represented by the small dashed line.
For (b) the dashed-dotted line shows the sum of $B$ background and continuum contributions.
}
\label{fig-eta}
\end{center}
\end{figure}

\begin{figure}[htb]
\begin{center}
\begin{minipage}{6.1in}
\includegraphics[width=\figwid]{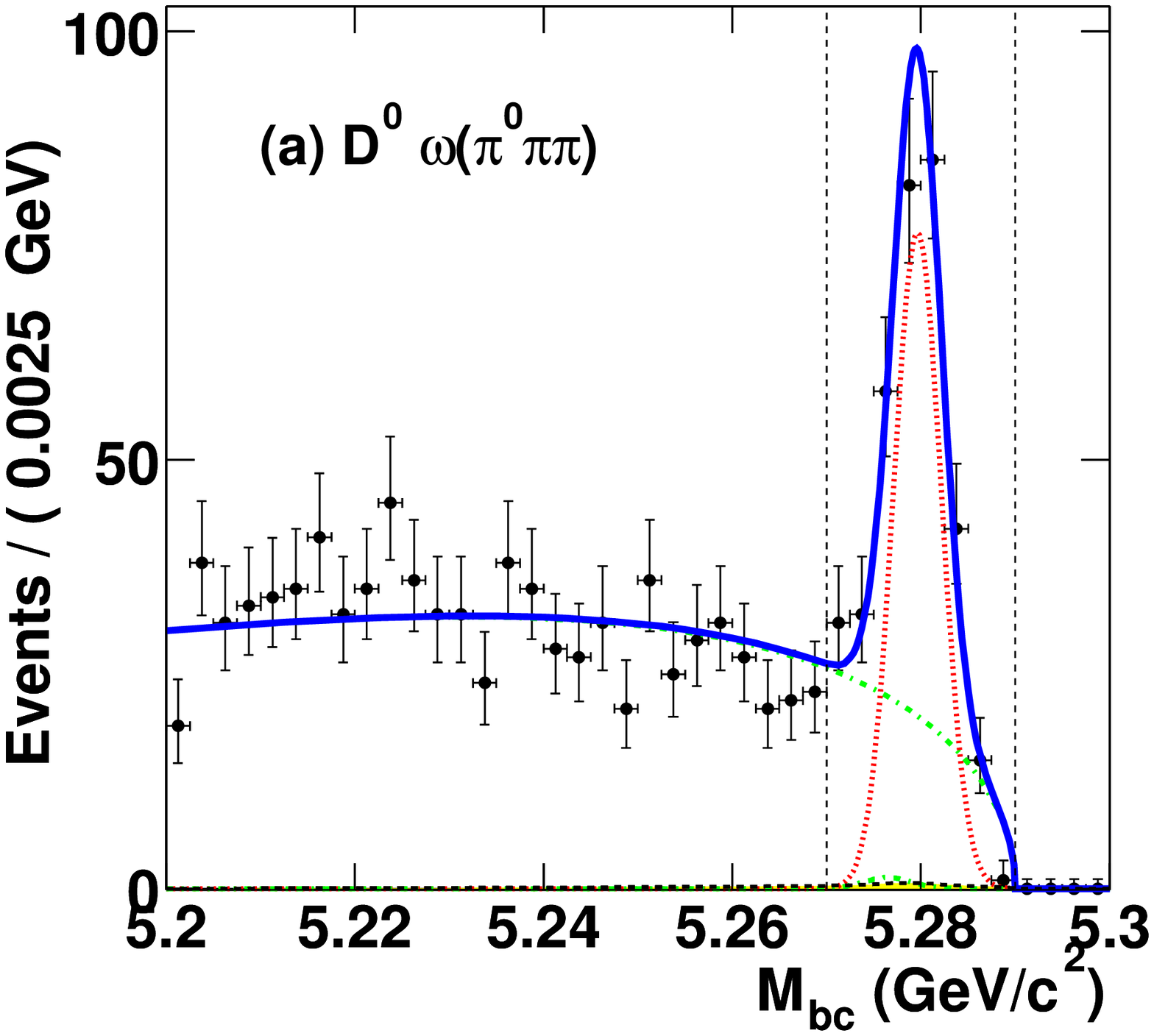}
\includegraphics[width=\figwid]{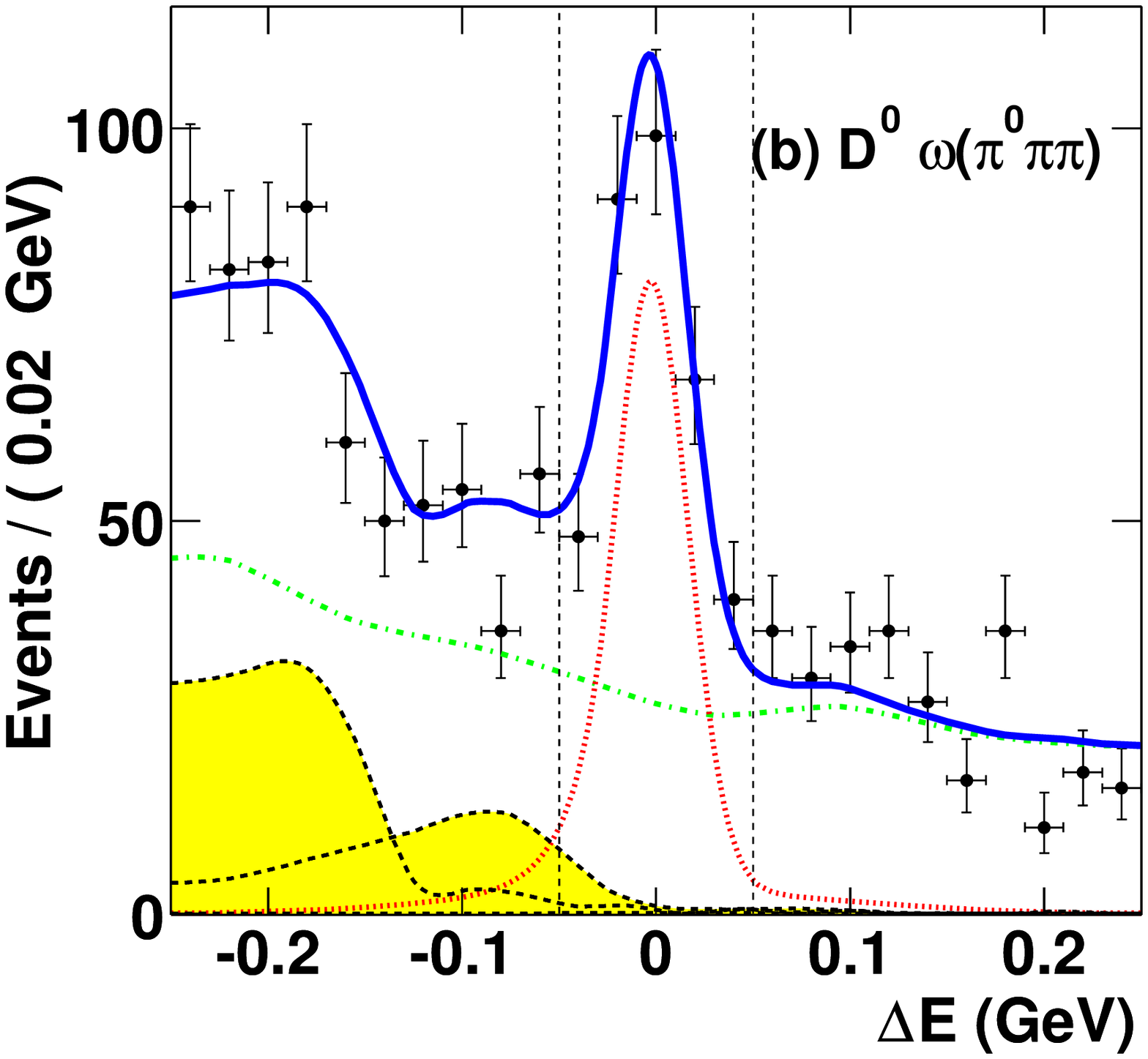}
\end{minipage}
\medskip
\caption{
Distributions of  (a) $\Mb$ and  (b) $\De$ for $\Bzerobar \to \Dzero \omega$ . Points with error bars 
represent the data, 
the solid line shows the result of the fit and the dotted line represents the signal contribution.
The vertical dashed lines represent the signal region.
For (a) the dashed-dotted line shows the continuum-like background contribution, with peaking background contributions
represented by the small dashed line.
For (b) the dashed-dotted lines show the sum of $B$ background and continuum contributions. 
}
\label{fig-omega}
\end{center}
\end{figure}

Significant background contributions arise both from
color favored decays 
and from contributions from other color suppressed decays (crossfeed) 
$\Bzerobar \to \Dstarzero \hzero $.  
Some backgrounds have the same final state
as the signal while others mimic signal due to missing or extra particles.

Generic Monte Carlo samples of $\BB$ and continuum $\qqbar$ are used  
to study the background contributions in the $\Mb$ and $\De$ distributions.
The $\BB$ event sample excludes the color suppressed modes under investigation and associated $\Dstarzero$ $\hzero$ modes. 
These signal modes for each of the decay chains considered and the corresponding $\Dstarzero$ $\hzero$ decays are
generated and reconstructed separately. They are used to estimate the crossfeed contributions to the other modes using the branching 
fractions measured here and by the BaBar collaboration~\cite{ref:Babar}. 
A combined generic Monte Carlo sample weighted according to the effective production cross sections 
and selection efficiencies of $\qqbar$ and $\BB$ is also used.

Figures~\ref{fig-pi0}, ~\ref{fig-eta} and ~\ref{fig-omega} 
show the $\Mb$  and $\De$ distributions after application of 
all selection requirements and with 
the $\De$ signal requirement applied for the $\Mb$ distributions (a) 
and
the signal requirement  $ 5.27 \GeVcsq  < \Mb < 5.29 \GeVcsq$ applied for the $\De$ distributions (b).
The $\De$ signal requirements used are  $|\De| < 0.08 \GeV $ for $\Dzero\pizero$ mode ,
$ -0.08 \GeV < \De < 0.05 \GeV $ for $\Dzero\eta$ ,  and $|\De| < 0.05 \GeV $ for $\Dzero\omega$. These regions are indicated 
by vertical dashed lines on the figures.

The dominant crossfeed contributions to the $\Dzero \hzero$ decays are found to arise from the corresponding
$\Dstarzero \hzero$ decays. These contributions peak at same $\Mb$ as the signal but are shifted to the lower side
in $\De$. As can be seen from the figures, the crossfeed contribution is substantial in the region $ -0.25 \GeV < \De < -0.10 \GeV$
but quite small in the signal region.    
Within the signal region the fraction of crossfeed is less than $10\%$ of the
observed yield in all cases. 

For the $\Bzerobar \to \Dzero\pizero$ mode the color allowed $\Bmi \to \Dstze \rhomi  $ modes are found to be the dominant backgrounds.
Non-reconstructed soft $\pizero$ from $\Dstarzero \to \Dzero \pizero $, photons from $\Dstarzero \to \Dzero \gamma $ and $\pimi$ 
from  $ \rhomi \to \pimi \pizero $  produce the same final state as the signal. 
However the missing particles cause a shift in $\De$ with a broad peak centered at approximately $\De = -0.2 \GeV$. 
In order to reduce contributions from this background, events that contain \B candidates
reconstructed as   $\Bmi \to \Dstze \rhomi $  within the signal region $ 5.27 \GeV < \Mb < 5.29 \GeV$ and $|\De| < 0.1 \GeV $ are rejected.
This requirement reduces the color allowed contribution in the region $-0.25 \GeV < \De < -0.10 \GeV$  by about $60 \%$ 
it does little to reduce contributions in the signal region, but remains useful to facilitate background modelling. 
The $\Mb$ distribution of these backgrounds is found to contribute at and slightly below the the $\Mb$ signal region.

\section{Data modelling and Signal Extraction}
Independent unbinned extended maximum likelihood fits to the $\Mb$ and $\De$ distributions are performed to obtain the signal yields. 
The yields from the $\Mb$ fits are used to extract the branching fractions, while the yields from the $\De$ 
fits are used to cross-check the results.
%
In most cases the shapes of the signal and background component distributions in $\Mb$ and $\De$ are obtained 
from fits to MC samples.
%
 
The signal models used are the same for all modes,  with the $\Mb$ signals modelled with a Gaussian function
 and the $\De$ signals with an empirical formula that accounts for the asymmetric calorimeter 
 energy response, known as the Crystal ball line shape~\cite{cbline}, added 
 to a Gaussian function of the same mean. The signal models are represented in 
Figures~\ref{fig-pi0}-~\ref{fig-omega} by the dotted lines.
Fits of the $\De$ distributions to signal Monte Carlo for each final state are used to obtain the 
signal shape parameters; all the $\Mb$ signal shape parameters are allowed to float in fits to data.

 The crossfeed contributions in $\Mb$ and $\De$ are studied using a combination of signal Monte Carlo samples from
 all other color suppressed modes, weighted according to the branching fractions obtained here or from the Babar 
 measurements~\cite{ref:Babar}. Smoothed histograms obtained from this combined sample are used as estimates of the  crossfeed 
 contributions.  In the $\Mb$ case the $\De$ signal region requirement results in very small crossfeed contributions, which
 are fixed at the Monte Carlo expectation. For $\De$ there are considerable contributions 
in the region  $ -0.25 \GeV < \De < -0.10 \GeV$; the normalization of this component is allowed to float in the fit.  
 

 Continuum like backgrounds in the $\Mb$ fits are modelled by an empirical threshold function known as 
 the ARGUS function~\cite{argus}.  The small peaking background contributions 
 are modelled by a Gaussian of mean and width and normalization obtained by a fit to the $B\bar{B}$ background Monte Carlo $\Mb$      distribution,  using an ARGUS function plus a Gaussian. A systematic uncertainty of 50\% is assigned to the determination
 of this small background distribution.   
 This treatment allows the vast majority of the background to be simply modelled with the ARGUS shape leaving a 
 small but less well known peaking background component  that represents the deviation from the ARGUS shape. 
 
 In fits to data $\Mb$ distributions the ARGUS background function parameters 
 are fixed to the values obtained from fits to combined Monte Carlo
 $B\bar{B}$ and continuum background samples.  The signal parameters are free, as are the 
 normalizations of signal and background. 
 The small peaking background and crossfeed contributions are fixed at their expected values.
  
  
The $\De$ background distributions in the $\Bzerobar \to \Dzero\eta$ 
and  $\Bzerobar \to \Dzero\omega$ are modelled  using smoothed histograms obtained from a combined continuum and
generic $\BB$ Monte Carlo sample.   

For the $\Bzerobar \to \Dzero\pizero$ mode, the shapes of the $\De$ distribution arising 
from $\BB$ and continuum background  are very different, necessitating separate modelling.
The continuum shape is modelled with a first order polynomial with slope obtained from fits to the continuum
Monte Carlo sample. The shape of the $\BB$ background is modelled with   
a Gaussian function plus a second order polynomial, with parameters determined from a fit to  
the generic $\BB$ Monte Carlo sample. In fits to data the large peak in the region $ -0.25 \GeV < \De < -0.10 \GeV $ that
arises principally from the color allowed $\Bmi \to \Dstze \rhomi $ decays is found to be broader than the Monte Carlo
expectation, thus all parameters of this color allowed Gaussian are allowed to float in the fit. The normalizations of the contributions 
from the remainder of the $\BB$ background, the continuum and the signal are also floated in the fit, with the small crossfeed
fixed as discussed above. 

The results of the $\Mb$ and $\De$ fits for the combined modes are presented in 
Figures~\ref{fig-pi0}, ~\ref{fig-eta} and ~\ref{fig-omega}. 

\begin{center}
\begin{table}[htbp]
\caption{ Measured signal region yields and MC estimates of signal region contributions for $\Bzerobar \to \Dzero \hzero$  for the combined $\Dzero$ subdecay modes and the individual $\Dzero$ subdecay modes.    The numbers of signal events $N_{sig}$ and continuum like events $N_{bkg}$ obtained from the $\Mb$ fit are listed together  with their statistical uncertainties.  MC estimates of the contributions from peaking $B$ background $N_{pkb}$ and crossfeeds from other color suppressed  modes $N_{xrs}$ are listed together with their systematic uncertainties. \newline }
\begin{tabular}{{@{\hspace{0.5cm}}l@{\hspace{0.5cm}}c@{\hspace{0.5cm}}c@{\hspace{0.5cm}}c@{\hspace{0.5cm}}c@{\hspace{0.5cm}}}}\hline\hline 
Mode & $N_{sig}$ & $N_{bkg}$ & $N_{pkb}$ & $N_{xrs}$ \\ 
\hline
$                           D^{0} \pi^{0}$ & $  637.1 \pm    33.9$ & $  613.4 \pm    25.6$ & $   21.1 \pm    10.5$ & $   25.9 \pm     6.5$ \\ 
$                     D^{0}(K\pi) \pi^{0}$ & $  216.6 \pm    17.1$ & $   94.3 \pm    10.2$ & $    6.1 \pm     3.0$ & $    6.9 \pm     1.7$ \\ 
$              D^{0}(K\pi\pi^{0}) \pi^{0}$ & $  184.5 \pm    17.6$ & $  191.9 \pm    14.0$ & $    4.8 \pm     2.4$ & $    7.6 \pm     1.9$ \\ 
$               D^{0}(K\pi\pi\pi) \pi^{0}$ & $  260.0 \pm    21.7$ & $  322.1 \pm    18.6$ & $   13.2 \pm     6.6$ & $   11.5 \pm     2.9$ \\ 
\hline
$                D^{0} \eta(\gamma\gamma)$ & $  161.3 \pm    16.8$ & $  135.6 \pm    12.6$ & $    4.5 \pm     2.2$ & $   10.7 \pm     2.7$ \\ 
$          D^{0}(K\pi) \eta(\gamma\gamma)$ & $   59.8 \pm     8.9$ & $   21.0 \pm     4.8$ & $    0.0 \pm     0.0$ & $    3.7 \pm     0.9$ \\ 
$   D^{0}(K\pi\pi^{0}) \eta(\gamma\gamma)$ & $   51.7 \pm     8.7$ & $   38.4 \pm     6.3$ & $    4.1 \pm     2.1$ & $    2.9 \pm     0.7$ \\ 
$    D^{0}(K\pi\pi\pi) \eta(\gamma\gamma)$ & $   70.6 \pm    10.7$ & $   72.5 \pm     9.4$ & $    2.5 \pm     1.3$ & $    4.1 \pm     1.0$ \\ 
\hline
$               D^{0} \eta(\pi^{0}\pi\pi)$ & $   68.5 \pm    10.6$ & $   41.6 \pm     6.3$ & $    1.2 \pm     0.6$ & $    2.0 \pm     0.5$ \\ 
$         D^{0}(K\pi) \eta(\pi^{0}\pi\pi)$ & $   21.4 \pm     5.3$ & $    7.2 \pm     2.6$ & $    0.8 \pm     0.4$ & $    0.5 \pm     0.1$ \\ 
$  D^{0}(K\pi\pi^{0}) \eta(\pi^{0}\pi\pi)$ & $   16.3 \pm     5.1$ & $   12.6 \pm     3.5$ & $    1.4 \pm     0.7$ & $    0.8 \pm     0.2$ \\ 
$   D^{0}(K\pi\pi\pi) \eta(\pi^{0}\pi\pi)$ & $   31.7 \pm     6.8$ & $   21.6 \pm     4.5$ & $    0.4 \pm     0.2$ & $    0.7 \pm     0.2$ \\ 
\hline
$                              D^{0} \eta$ & $  233.1 \pm    19.9$ & $  173.9 \pm    13.8$ & $    5.5 \pm     2.7$ & $   12.7 \pm     3.2$ \\ 
\hline
$             D^{0} \omega(\pi^{0}\pi\pi)$ & $  191.9 \pm    17.8$ & $  154.1 \pm    12.1$ & $    3.4 \pm     1.7$ & $    3.5 \pm     0.9$ \\ 
$       D^{0}(K\pi) \omega(\pi^{0}\pi\pi)$ & $   73.5 \pm     9.8$ & $   31.8 \pm     5.7$ & $    3.0 \pm     1.5$ & $    0.8 \pm     0.2$ \\ 
$D^{0}(K\pi\pi^{0}) \omega(\pi^{0}\pi\pi)$ & $   54.0 \pm     8.9$ & $   43.5 \pm     6.2$ & $    0.0 \pm     0.0$ & $    1.3 \pm     0.3$ \\ 
$ D^{0}(K\pi\pi\pi) \omega(\pi^{0}\pi\pi)$ & $   73.2 \pm    11.0$ & $   77.8 \pm     8.5$ & $    1.0 \pm     0.5$ & $    1.4 \pm     0.3$ \\ 
\hline\hline
\end{tabular}
\label{yield-mb-sub}
\end{table}
\end{center}

\section{Branching Fraction results} 

\begin{center}
\begin{table}[htbp]
\caption{Efficiency from Monte Carlo $\epsilon_{MC} $, correction factor $\epsilon_{DA/MC}$, and   corrected efficiency $\epsilon_{corr} $  for the combined modes and the individual submodes.  This efficiency is for the 1d $\Mb$ fit sample, with the $\De$ signal region requirement applied. The relative uncertainty is given in brackets. \newline}
\begin{tabular}{{@{\hspace{0.5cm}}l@{\hspace{0.5cm}}c@{\hspace{0.5cm}}c@{\hspace{0.5cm}}c@{\hspace{0.5cm}}}}\hline\hline 
Mode & $\epsilon_{MC}  $ &    $\epsilon_{DA/MC}$ & $\epsilon_{corr} $ \\ 
\hline
$                           D^{0} \pi^{0}$ & $  0.078 \pm   0.002 (1.9\%) $ & $  0.938 \pm   0.048 (5.1\%) $ & $  0.075 \pm   0.004 (5.5\%) $ \\ 
$                     D^{0}(K\pi) \pi^{0}$ & $  0.177 \pm   0.002 (1.4\%) $ & $  0.979 \pm   0.045 (4.6\%) $ & $  0.173 \pm   0.008 (4.8\%) $ \\ 
$              D^{0}(K\pi\pi^{0}) \pi^{0}$ & $  0.047 \pm   0.001 (2.9\%) $ & $  0.905 \pm   0.044 (4.9\%) $ & $  0.042 \pm   0.002 (5.7\%) $ \\ 
$               D^{0}(K\pi\pi\pi) \pi^{0}$ & $  0.087 \pm   0.001 (1.5\%) $ & $  0.979 \pm   0.057 (5.8\%) $ & $  0.085 \pm   0.005 (6.0\%) $ \\ 
\hline
$                D^{0} \eta(\gamma\gamma)$ & $  0.060 \pm   0.001 (2.0\%) $ & $  0.989 \pm   0.058 (5.9\%) $ & $  0.061 \pm   0.004 (6.2\%) $ \\ 
$          D^{0}(K\pi) \eta(\gamma\gamma)$ & $  0.134 \pm   0.002 (1.6\%) $ & $  1.032 \pm   0.057 (5.5\%) $ & $  0.139 \pm   0.008 (5.7\%) $ \\ 
$   D^{0}(K\pi\pi^{0}) \eta(\gamma\gamma)$ & $  0.038 \pm   0.001 (2.6\%) $ & $  0.954 \pm   0.054 (5.7\%) $ & $  0.036 \pm   0.002 (6.2\%) $ \\ 
$    D^{0}(K\pi\pi\pi) \eta(\gamma\gamma)$ & $  0.064 \pm   0.001 (1.8\%) $ & $  1.032 \pm   0.067 (6.5\%) $ & $  0.066 \pm   0.004 (6.7\%) $ \\ 
\hline
$               D^{0} \eta(\pi^{0}\pi\pi)$ & $  0.044 \pm   0.001 (2.5\%) $ & $  0.940 \pm   0.055 (5.9\%) $ & $  0.042 \pm   0.003 (6.4\%) $ \\ 
$         D^{0}(K\pi) \eta(\pi^{0}\pi\pi)$ & $  0.096 \pm   0.002 (1.8\%) $ & $  0.981 \pm   0.054 (5.5\%) $ & $  0.094 \pm   0.005 (5.8\%) $ \\ 
$  D^{0}(K\pi\pi^{0}) \eta(\pi^{0}\pi\pi)$ & $  0.028 \pm   0.001 (3.5\%) $ & $  0.906 \pm   0.052 (5.7\%) $ & $  0.025 \pm   0.002 (6.7\%) $ \\ 
$   D^{0}(K\pi\pi\pi) \eta(\pi^{0}\pi\pi)$ & $  0.046 \pm   0.001 (2.1\%) $ & $  0.981 \pm   0.064 (6.5\%) $ & $  0.046 \pm   0.003 (6.8\%) $ \\ 
\hline
$                              D^{0} \eta$ & $  0.054 \pm   0.001 (2.1\%) $ & $  0.971 \pm   0.057 (5.9\%) $ & $  0.054 \pm   0.003 (6.3\%) $ \\ 
\hline
$             D^{0} \omega(\pi^{0}\pi\pi)$ & $  0.028 \pm   0.001 (2.9\%) $ & $  0.885 \pm   0.061 (6.9\%) $ & $  0.025 \pm   0.002 (7.5\%) $ \\ 
$       D^{0}(K\pi) \omega(\pi^{0}\pi\pi)$ & $  0.060 \pm   0.001 (2.3\%) $ & $  0.924 \pm   0.061 (6.6\%) $ & $  0.055 \pm   0.004 (7.0\%) $ \\ 
$D^{0}(K\pi\pi^{0}) \omega(\pi^{0}\pi\pi)$ & $  0.017 \pm   0.001 (3.8\%) $ & $  0.854 \pm   0.058 (6.8\%) $ & $  0.015 \pm   0.001 (7.8\%) $ \\ 
$ D^{0}(K\pi\pi\pi) \omega(\pi^{0}\pi\pi)$ & $  0.032 \pm   0.001 (2.5\%) $ & $  0.924 \pm   0.069 (7.4\%) $ & $  0.029 \pm   0.002 (7.8\%) $ \\ 
\hline\hline
\end{tabular}
\label{eff-mb-sub}
\end{table}
\end{center}

\begin{center}
\begin{table}[htbp]
\caption{ Measured branching fractions for the process $\Bzerobar \to \Dzero \hzero$ $(\times 10^{-4})$   using separate $\Dzero$ subdecay mode samples,   as obtained from the $\Mb$ fit.   The branching fractions are listed with statistical and systematic uncertainties. \newline}
\begin{tabular}{{@{\hspace{0.5cm}}l@{\hspace{0.5cm}}c@{\hspace{0.5cm}}c@{\hspace{0.5cm}}c@{\hspace{0.5cm}}}}\hline\hline 
Mode  & $                            D^{0}(K\pi) $ & $                     D^{0}(K\pi\pi^{0}) $ & $                      D^{0}(K\pi\pi\pi) $ \\ \hline
$                           D^{0} \pi^{0}$ & $   2.20 \pm    0.17 \pm    0.18 $ & $   2.11 \pm    0.20 \pm    0.24 $ & $   2.75 \pm    0.23 \pm    0.27 $ \\ 
$                D^{0} \eta(\gamma\gamma)$ & $   1.89 \pm    0.28 \pm    0.18 $ & $   1.73 \pm    0.29 \pm    0.22 $ & $   2.39 \pm    0.36 \pm    0.26 $ \\ 
$               D^{0} \eta(\pi^{0}\pi\pi)$ & $   1.69 \pm    0.42 \pm    0.15 $ & $   1.32 \pm    0.41 \pm    0.17 $ & $   2.66 \pm    0.57 \pm    0.26 $ \\ 
$             D^{0} \omega(\pi^{0}\pi\pi)$ & $   2.55 \pm    0.34 \pm    0.27 $ & $   1.95 \pm    0.32 \pm    0.27 $ & $   2.45 \pm    0.37 \pm    0.29 $ \\ 
\hline\hline
\end{tabular}
\label{brr-mb-sub}
\end{table}
\end{center}

Results of $\Mb$ and $\De$ fits are consistent; the agreement is typically within 50\% of the statistical uncertainty. 
The results from the $\Mb$ fits are found to have a slightly smaller total uncertainty in most cases and are used for the final result.
Yields are obtained both from the individual subdecay mode samples and from samples with the three $\Dzero$ subdecay samples
combined. The yields from the one dimensional $\Mb$ fits are shown in Table~\ref{yield-mb-sub}. Both peaking backgrounds and crossfeed 
contributions in the signal region can be seen to contribute substantially less than the 
extent of the statistical uncertainty on the signal yield.


The yields obtained are interpreted as branching fractions using the number of analyzed $\BB$ events,
the product of subdecay fractions from PDG~\cite{pdg}  corresponding to the decay of  $\Dzero \hzero$ into the 
observed final states and the total selection efficiency. The efficiency for each mode is first obtained from signal Monte Carlo
samples and then corrected by  comparing data and MC predictions for other processes.
For the $\pizero$ reconstruction efficiency the correction is obtained from comparisons 
of $\eta \to \pizero \pizero \pizero $ to $\eta \to \gamgam$ and to $\eta \to \threepi$, for data and Monte Carlo.  
The MC efficiency, correction factor and corrected efficiency are presented in Table~\ref{eff-mb-sub}. 
The corrections are obtained from the product of correction factors relevant to the final state of each submode. 

The branching fraction results for 
the individual submodes are shown in Table~\ref{brr-mb-sub}.
The combined submode systematic uncertainties and branching fraction results are 
shown in Tables~\ref{tot-mb} and ~\ref{brr-mb}, respectively. 

\section{Systematic Uncertainties} 

\begin{center}
\begin{table}[htbp]
\caption{Systematic uncertainties of the measured branching fractions for $\Bzerobar \to \Dzero \hzero $,    for the combined $\Dzero$ submode samples,   as estimated for the $\Mb$ fit results.   \newline}
\begin{tabular}{lccccc}\hline\hline 
Category & $                           D^{0} \pi^{0}$ & $                D^{0} \eta(\gamma\gamma)$ & $               D^{0} \eta(\pi^{0}\pi\pi)$ & $                              D^{0} \eta$ & $             D^{0} \omega(\pi^{0}\pi\pi)$ \\ 
\hline
Tracking efficiency & 2.6 & 2.6 & 2.6 & 2.6 & 2.6 \\ 
$\hzero$ efficiency & 2.7 & 4.0 & 4.0 & 4.0 & 5.4 \\ 
Kaon efficiency & 1.0 & 1.0 & 1.0 & 1.0 & 1.0 \\ 
Extra $\pizero$ efficiency  & 0.4 & 0.4 & 0.4 & 0.4 & 0.4 \\ 
Likelihood ratio efficiency  & 3.0 & 3.0 & 3.0 & 3.0 & 3.0 \\ 
MC statistics   & 2.3 & 2.2 & 2.8 & 2.4 & 3.2 \\ 
Peaking background & 1.7 & 1.4 & 0.9 & 1.2 & 0.9 \\ 
Crossfeed & 1.0 & 1.7 & 0.7 & 1.4 & 0.5 \\ 
$\Delta E$ resolution & 5.0 & 5.0 & 5.0 & 5.0 & 5.0 \\ 
Modelling $\pm1\sigma$ variations & 2.0 & 5.4 & 0.2 & 10.6 & 3.1 \\ 
Branching Fractions $D^{0}$, $\pi^{0}$, $\eta$, $\omega$  & 5.2 & 5.2 & 5.5 & 5.3 & 5.2 \\ 
Number of $B\bar{B}$ events & 0.7 & 0.7 & 0.7 & 0.7 & 0.7 \\ 
\hline
Total (\%) & 9.5 & 11.1 & 9.9 & 14.4 & 11.0 \\ 
\hline\hline
\end{tabular}
\label{tot-mb}
\end{table}
\end{center}

%
Systematic uncertainties of the combined modes, estimated for the results based 
on the $\Mb$ fits are summarized in Table~\ref{tot-mb}.
Uncertainties on the efficiency correction factors relevant to each final state are listed in the Tables 
together with other uncertainties.  For the $\Mb$ fit the uncertainty from the peaking background, which is fixed at the MC expectation in the fit,
is obtained by propagating a 50\% uncertainty on the normalization of this contribution. The crossfeed uncertainty is estimated as 25\% of the
contribution from this source in the signal regions. This accounts for uncertainties on the branching fractions of the crossfeed contributions
and also differences observed between the floated crossfeed contributions in $\De$ fits and the MC expectation.     
Uncertainties arising from the background and signal modelling used are estimated from the changes in the yields as a 
result of $\pm 1 \sigma$ variations on the model parameters. 
The total uncertainty is obtained regarding uncertainties from different sources as uncorrelated.  

\begin{center}
\begin{table}[htbp]
\caption{ Measured branching fractions for the process $\Bzerobar \to \Dzero \hzero$ $(\times 10^{-4})$   using combined $\Dzero$ subdecay mode samples,   as obtained from the $\Mb$ fit.   The branching fractions are listed with statistical and systematic uncertainties.  The $\Dzero \eta $ result is obtained from a combined sample of $\Dzero \eta(\gamgam) $ and $\Dzero \eta(\threepi) $.  \newline}
\begin{tabular}{lc}\hline\hline 
Mode  & Branching fraction $( \times 10^{-4} )$ \\ \hline
$                           D^{0} \pi^{0}$ & $   2.31 \pm    0.12 \pm    0.23 $ \\ 
$                D^{0} \eta(\gamma\gamma)$ & $   1.77 \pm    0.18 \pm    0.20 $ \\ 
$               D^{0} \eta(\pi^{0}\pi\pi)$ & $   1.89 \pm    0.29 \pm    0.20 $ \\ 
$                              D^{0} \eta$ & $   1.83 \pm    0.15 \pm    0.27 $ \\ 
$             D^{0} \omega(\pi^{0}\pi\pi)$ & $   2.25 \pm    0.21 \pm    0.28 $ \\ 
\hline\hline
\end{tabular}
\label{brr-mb}
\end{table}
\end{center}

\section{Conclusion}

Improved measurements of the branching fractions of the color-suppressed decays  $\Bzerobar \to \Dzero \pizero$, 
$ \Dzero \eta $ and $ \Dzero \omega $ are presented. 
The results are consistent with the previous Belle measurements.
The total uncertainty of the new results is two to three times smaller than the previous results,
 mostly due to the seven times larger data sample.
However comparing the results with those of BaBar~\cite{ref:Babar} and CLEO~\cite{ref:CLEO} indicates an approximately
2$\sigma$ difference, with all three branching fractions measured here lower than the previous measurements. 
    
All the branching fraction results are similar, in the range 1.8-2.4 $\times 10^{-4}$.
The large values disfavour theoretical predictions based on naive factorization descriptions
and indicate the need for models including final state interaction effects to satisfactorily  
describe the observations.

%
%
\clearpage

\section*{Acknowledgments}
We thank the KEKB group for the excellent operation of the
accelerator, the KEK Cryogenics group for the efficient
operation of the solenoid, and the KEK computer group and
the National Institute of Informatics for valuable computing
and Super-SINET network support. We acknowledge support from
the Ministry of Education, Culture, Sports, Science, and
Technology of Japan and the Japan Society for the Promotion
of Science; the Australian Research Council and the
Australian Department of Education, Science and Training;
the National Science Foundation of China under contract
No.~10175071; the Department of Science and Technology of
India; the BK21 program of the Ministry of Education of
Korea and the CHEP SRC program of the Korea Science and
Engineering Foundation; the Polish State Committee for
Scientific Research under contract No.~2P03B 01324; the
Ministry of Science and Technology of the Russian
Federation; the Ministry of Education, Science and Sport of
the Republic of Slovenia; the National Science Council and
the Ministry of Education of Taiwan; and the U.S.\
Department of Energy.


\begin{thebibliography}{99}
\bibitem{CC}
Throughout this paper, 
the inclusion of the charge conjugate mode decay is implied
unless otherwise stated.
\end{thebibliography}
\end{document}